\documentstyle[12pt]{article}

\begin{document}
\begin{titlepage}
\begin{centering}
\vspace{4.0cm}
{\LARGE{\bf Radial fingering in a Hele-Shaw cell: 
a weakly nonlinear analysis}}\\
\bigskip\bigskip
\vspace{3.0cm}
Jos\'e A. Miranda\footnote[1]{e-mail:01jamn@npd.ufpe.br} and Michael Widom\footnote[2]{e-mail:widom@andrew.cmu.edu}\\
{\em Department of Physics\\
Carnegie Mellon University\\
Pittsburgh, PA 15213\\  
USA}\\

\vspace{1.0cm}
\end{centering}
\pagebreak

\begin{abstract}
The Saffman-Taylor viscous fingering instability occurs when a less
viscous fluid displaces a more viscous one between narrowly spaced
parallel plates in a Hele-Shaw cell. Experiments in radial flow
geometry form fan-like patterns, in which fingers of different lengths
compete, spread and split.  Our weakly nonlinear analysis of the
instability predicts these phenomena, which are beyond the scope of
linear stability theory.  Finger competition arises through enhanced
growth of sub-harmonic perturbations, while spreading and splitting
occur through the growth of harmonic modes. Nonlinear mode-coupling
enhances the growth of these specific perturbations with appropriate
relative phases, as we demonstrate through a symmetry analysis of the
mode coupling equations. We contrast mode coupling in radial flow with
rectangular flow geometry.
\end{abstract}
\begin{center}
\end{center}
\begin{center}
\end{center}
\end{titlepage}
\def\carre{\vbox{\hrule\hbox{\vrule\kern 3pt
\vbox{\kern 3pt\kern 3pt}\kern 3pt\vrule}\hrule}}

\baselineskip = 30pt
\section{Introduction}

The Saffman-Taylor problem~\cite{Rev} addresses motion of two viscous
fluids in the narrow space between two plates, known as a Hele-Shaw
cell.  When a fluid of low viscosity displaces a fluid of higher
viscosity, the interface between them becomes unstable and starts to
deform.  This problem is formally equivalent~\cite{BenJac-2} to a
special case of a solidification problem.  Experiments and theory
focus on two basic flow geometries (i) rectangular~\cite{Saf} and (ii)
radial~\cite{Pat}.  For both geometries, the initial developments of
the interface instability tracks the predictions of linear stability
theory~\cite{Rev}.  As the unstable modes of perturbation grow, they
become coupled in a weakly nonlinear stage of evolution. Finally, the
system evolves to a complicated late stage, characterized by formation
of fingering structures, in which nonlinear effects dominate.

This paper addresses linear and the weakly nonlinear stages of the
evolution. Here we concentrate our attention on the radial flow
geometry. We develop the mode coupling theory for the Saffman-Taylor
problem.  Our approach is a complementary study to the purely linear
investigations~\cite{Rev} and to recent developments in the
analytical~\cite{Tan} and numerical~\cite{Hou} treatments of the fully
nonlinear time-evolving flow. These analytical works~\cite{Tan}
describe the early and late stage dynamics of the interface in the
small surface tension limit, using conformal mapping techniques.
Efficient numerical simulations~\cite{Hou} describe complex flow
patterns, for both large and small surface tension, employing a
boundary integral formulation of the equations of motion. On the other
hand, our analytical weakly nonlinear study investigates the
intermediate stage between the purely linear and the fully nonlinear
ones, focusing on the onset of the nonlinear effects. Our approach
applies to any value of surface tension, and gives physical insight
into mechanisms of pattern solution. In the context of solidification,
both numerical~\cite{Brush2} and weakly nonlinear~\cite{Brush,Debroy}
results are known.

In a separate publication~\cite{Mir} we develop 
the weakly nonlinear theory for rectangular flow geometry. 
Although mathematically similar to the rectangular 
flow problem, the radial geometry imposes two important new elements: 
(i) the interface is always asymmetric because of the distinction 
between the area inside and outside of the boundary between the 
fluids; (ii) the radial flow presents multiple stages of instability, 
with no true steady state attainable. 

Experimental and theoretical investigations are plentiful for radial 
flow geometry Hele-Shaw cells~\cite{Rev}. These 
experiments~\cite{{Pat},{Che},{Tho},{Bat}} inject the less viscous 
fluid at the center 
of the cell, which was previously filled with the more viscous fluid. 
The injection is usually performed at a constant flow rate or 
constant pressure. Initially, the interface (bubble) is circular. 
As the bubble develops, the interface undulates and fingers grow. 
As the size of the bubble increases, the fingers spread and their tips 
become blunt, creating a fan-like pattern. Next, 
the fingers start to split at the tip. Larger fingers 
shield smaller ones and the pattern develops asymmetrically, causing 
competition among fingers of different lengths. Spreading, splitting and 
competition are the three basic growth mechanisms of the viscous 
fingering process~\cite{Hom}.

The linear stability of a circular interface evolving in a radial 
geometry flow Hele-Shaw cell was studied by Bataille~\cite{Bat}, 
Wilson~\cite{Wil}, and thereafter by Paterson~\cite{Pat}. More recent 
experimental and theoretical work study the linear stage of the 
radial Hele-Shaw flow in a variety of situations: Sader, Chan and 
Hughes~\cite{Sad} examined the linear stability analysis for the case 
of a non-Newtonian displaced fluid; Carrillo et al.~\cite{Car2} considered 
radial flow in a rotating Hele-Shaw cell; Cardoso and Woods~\cite{Car} 
investigated the linear stability analysis of an annulus of fluid, bound 
by two other fluids of different viscosities. 
Because of the complexity of the 
evolution, even at the level of linear analysis, this paper reviews 
certain results of the linear theory for the sake of comparison 
with the nonlinear analysis.

The linear stability analysis applies only to early stages of the
flow. Moreover, it does not explain well the essential mechanisms of
spreading, splitting and competition cited above.  To analyze these
effects, we develop the weakly nonlinear mode coupling theory of the
radial flow in a Hele-Shaw cell.  We focus on the onset of the
nonlinear effects and their influence on the evolution of the unstable
interface.  We employ an analytical approach known as a mode coupling
theory.  The paper is organized as follows: section II derives a
differential equation describing the early nonlinear evolution of the
interface modes. In section III, we interpret results obtained in
section II and investigate both the linear and weakly nonlinear
evolution of the system. Our chief conclusions are presented in this
section. We show that second order mode coupling explains the main
features of patterning formation by coupling fundamental modes to
harmonics and sub-harmonics. Symmetries of the radial flow geometry
dictate the form of the mode coupling equations, influencing the
shapes of growing interfaces. Section IV presents our final remarks.

\section{Derivation of the mode coupling differential equation} 
\label{derivation}

Consider two immiscible, incompressible, viscous fluids. In a Hele-Shaw cell 
fluids flow in a narrow gap of thickness $b$, between two parallel 
plates (see figure 1). We assume that $b$ is smaller than any other 
length scale in the problem, and therefore the system is considered to 
be effectively two-dimensional. Denote the viscosities of the inner 
and outer fluids, respectively as $\eta_{1}$ and $\eta_{2}$. 
The flows are assumed to be irrotational, except at the interface. 
Inject fluid 1 into fluid 2 at constant flow rate $Q$, equal to the 
area covered per unit time. Between the two fluids there exists 
a surface tension $\sigma$.

During the flow, the interface has a perturbed shape described as 
${\cal R}= R + \zeta(\theta,t)$, where $\theta$ represents the polar 
angle, and $R=R(t)$ denotes the time dependent unperturbed radius 
\begin{equation}
\label{R}
R(t)=\sqrt{R_{0}^{2} + \frac{Qt}{\pi}}.
\end{equation}
$R_{0}$ is the unperturbed radius at $t=0$.
The analytic model we seek 
predicts the evolution of the interface perturbation amplitude 
$\zeta(\theta,t)$. The early nonlinear evolution of the interface obeys a
second-order mode coupling equation.

The linear evolution is most readily described in terms of 
the complex Fourier modes
\begin{equation}
\zeta_{n}(t) = \frac{1}{2\pi} \int_{0}^{2\pi} \zeta(\theta,t) \exp{(-in\theta)} {\rm d}\theta,
\end{equation}
where $n$=0,$\pm 1$, $\pm 2$, $...$ denotes the discrete azimuthal wave number.
We remain in Fourier space for our weakly nonlinear analysis. 
The net perturbation is represented by
\begin{equation}
\label{z}
\zeta(\theta,t)=\sum_{n=-\infty}^{+\infty} \zeta_{n}(t) \exp{(i n \theta)}. 
\end{equation}
Note that we include 
the $n=0$ mode in our Fourier expansion~(\ref{z}). This is done to 
keep the area ${\cal A}=\pi R^{2}=\pi R_{0}^{2} + Qt$ 
of the perturbed shape independent of 
the perturbation $\zeta$.
Thus we require
\begin{equation}
\label{z0}
\zeta_{0}= -\frac{1}{2R}\sum_{n=1}^{\infty} \left [ |\zeta_{n}(t)|^{2} + |\zeta_{-n}(t)|^2 \right ].
\end{equation}
The constant area constraint is 
intrinsically a nonlinear concern and is not required in linear stability 
analysis. Since we are interested in the early nonlinear behavior of 
the system, our first task is to derive a differential equation 
for $\zeta_{n}$, correct to second order.

We will occasionally refer to the ``rectangular geometry limit''. By 
this we mean the limit $R \rightarrow \infty$ and $Q \rightarrow \infty$ 
such that $Q/(2 \pi R) \equiv 
v_{\infty}$ and $n/R \equiv k$ remain constant, where $v_{\infty}$ is 
the flow velocity at infinity and $k$ denotes the wave number 
of the disturbance. In this limit the interface evolution reverts 
to the simpler evolution of the rectangular flow geometry. 
For example, the constant area constraint 
becomes simple ($\zeta_{0}=0$) in this limit.

The relevant hydrodynamic equation is Darcy's law~\cite{Rev,Saf}
\begin{equation}
\label{Darcy}
-\frac{12\eta_{i}}{b^{2}}\vec v_{i} - \vec\nabla p_{i} = 0,
\end{equation}
where $\vec v_{i}=\vec v_{i}(r,\theta)$ 
and $p_{i}=p_{i}(r,\theta)$ are, 
respectively, the velocity and pressure in fluids 
$i=1$ and $2$. At the interface, the pressure difference 
between the two fluids is given 
by~\cite{Rev}
\begin{equation}
\label{pressure}
(p_{1} - p_{2})|_{{\cal R}} = \sigma \left ( \kappa_{\|} + \kappa_{\bot} \right )|_{{\cal R}}.
\end{equation}
The curvature in the direction perpendicular to the plates ($\kappa_{\bot}$) 
is nearly constant~\cite{McL,Par}. Because its gradient is nearly zero, it 
does not significantly affect the motion in our problem.

Due to the radial geometry, we have a somewhat complicated expression for the 
in-plane interface curvature~\cite{Wil,Dub}
\begin{equation}
\label{parallel}
\kappa_{\|}=\frac{ \left [ r^2 + 2 \left ( \frac{\partial r}{\partial \theta} \right )^2 - r \frac{\partial^2 r}{\partial \theta^{2}} \right ]}{ \left [ r^2 + \left (\frac{\partial r}{\partial \theta} \right )^2 \right ]^{3/2}},
\end{equation}
where $r$ denotes the distance from the radial flow source.
The sign convention for the curvature $\kappa_{\|}$ is such that a 
circular interface has positive curvature. Keeping second order terms 
in the perturbation amplitude $\zeta$, we rewrite the in-plane 
curvature
\begin{equation}
\label{kappa}
\kappa_{\|}=\left \{ \frac{1}{R} - \frac{1}{R^{2}} \left ( \zeta + \frac{\partial^2 \zeta}{\partial \theta^{2}} \right ) + \frac{1}{R^{3}} \left [ \zeta^{2} + \frac{1}{2}\left ( \frac{\partial \zeta}{\partial \theta} \right )^2 + 2 \zeta \frac{\partial^2 \zeta}{\partial \theta^{2}} \right ] \right \}.
\end{equation}
For future reference, note that only one term 
$(1/R^{2}) (\partial^2 \zeta/ \partial \theta^{2})$ out of six terms in equation~(\ref{kappa}) survives in the rectangular 
geometry limit.

Taking advantage of the irrotational and incompressible flow conditions, 
we define the velocity potential $\phi_{i}$ in each of the fluids, 
where $\vec v_{i}=-\vec \nabla\phi_{i}$. The velocity potential 
satisfies Laplace's equation $\nabla^{2}\phi_{i}=0$. Combining 
the velocity potential with equations~(\ref{pressure}) 
and~(\ref{parallel}) for the pressure difference and Darcy's law~(\ref{Darcy}),
we write the equation of motion
\begin{equation}
\label{dimensionless2}
A \left ( \frac{\phi_{1}|_{{\cal R}} + \phi_{2}|_{{\cal R}}}{2} \right ) -  \left ( \frac{\phi_{1}|_{{\cal R}} - \phi_{2}|_{{\cal R}}}{2} \right ) = - \alpha \kappa_{\|},
\end{equation}
where
\begin{equation}
\label{contrast}
A=\frac{\eta_{2} - \eta_{1}}{\eta_{2} + \eta_{1}}
\end{equation}
is the viscosity contrast and 
\begin{equation} 
\label{alpha}
\alpha=\frac{b^{2} \sigma}{12(\eta_{1} + \eta_{2})}.
\end{equation}
Equation~(\ref{dimensionless2}) drops an arbitrary constant of integration
related to the definition of velocity potentials.
For our weakly nonlinear analysis we are 
interested in second order contributions in the perturbation amplitudes.
Therefore, all the quantities in equation~(\ref{dimensionless2}) are 
evaluated at the perturbed interface ${\cal R}=R + \zeta(\theta,t)$, and
not at the initial unperturbed interface position $R$ as is usually done in 
linearized surface deformation problems. The 
nonlinear terms arise because of this important distinction.

Now define Fourier expansions for the velocity
potentials $\phi_{i}$. 
Far from the interface the velocity field 
should approach the unperturbed steady flow with a circular interface 
of radius R. Thus for $ r \rightarrow 0$ and $r \rightarrow +\infty$ 
the velocity potentials $\phi_{i}$ approach $\phi_{i}^{0}$, 
the velocity potentials for radial flow
\begin{equation}
\label{steady}
\phi_{i}^{0}=- \frac{Q}{2 \pi} \log{ \left (\frac{r}{R} \right )} + C_{i}.
\end{equation}
$C_{i}$ are independent of $r$ and $\theta$. They do not contribute to the 
velocity fields $\vec v_{i}$, so we will not discuss them further.
The general velocity potentials obeying these requirements are 
\begin{equation}
\label{phi1}
\phi_{1}= \phi_{1}^{0} + \sum_{n \neq 0} \phi_{1 n}(t) \left ( \frac{r}{R} \right )^{|n|} \exp(i n \theta),
\end{equation}
and
\begin{equation}
\label{phi2}
\phi_{2}= \phi_{2}^{0} + \sum_{n \neq 0} \phi_{2 n}(t) \left ( \frac{R}{r} \right )^{|n|} \exp(i n \theta).
\end{equation}
The algebraic dependence on $r$ transforms, in the rectangular geometry 
limit, into exponential dependence.
In order to calculate the mode coupling differential 
equation for the system, we 
substitute expansions~(\ref{steady}),~(\ref{phi1}) and ~(\ref{phi2}) into 
the equation of motion~(\ref{dimensionless2}), keep second order terms 
in the perturbation amplitudes, and Fourier transform them.

We need additional relations expressing 
the velocity potentials in terms of the perturbation amplitudes. 
To find these, consider 
the kinematic boundary condition which states that
the normal components of each fluid's velocity at the interface equals 
the velocity of the interface itself~\cite{Ros}. In polar coordinates $(r, \theta)$ the kinematic boundary condition is written as
\begin{equation}
\label{b.c.}
\frac{\partial {\cal R}}{\partial t}= \left ( \frac{1}{r^2} \frac{\partial {\cal R}}{\partial \theta}\frac{\partial \phi_{i}}{\partial \theta} \right )_{r={\cal R}} - \left (\frac{\partial \phi_{i}}{\partial r} \right )_{r={\cal R}}.
\end{equation}
Expand the boundary condition equation~(\ref{b.c.}) to second order 
in $\zeta$ and then Fourier transform. The constraint that area ${\cal A}$ 
be independent of the perturbation $\zeta$ (equation~(\ref{z0})) enforces 
the equality of the $n=0$ Fourier mode on each side of equation~(\ref{b.c.}). 
Nontrivial identities are obtained for $n \neq 0$.

Solving for $\phi_{in}(t)$ 
consistently to second order in $\zeta$ yields
\begin{eqnarray}
\label{phi1t}
\phi_{1n}(t) & = & -\frac{R}{|n|}\dot{\zeta}_{n} - \frac{Q}{2 \pi R |n|} \zeta_{n} \nonumber \\
             & + & \sum_{n' \neq 0} \left ( sgn(nn') - \frac{1}{|n|} \right ) \dot{\zeta}_{n'}\zeta_{n - n'} + \frac{Q}{2 \pi R^{2}} \sum_{n' \neq 0} sgn(nn') \zeta_{n'}\zeta_{n - n'}, \nonumber \\
\end{eqnarray}
and
\begin{eqnarray}
\label{phi2t} 
\phi_{2n}(t) & = & \frac{R}{|n|}\dot{\zeta}_{n} + \frac{Q}{2 \pi R |n|} \zeta_{n} \nonumber \\
             & + & \sum_{n' \neq 0} \left ( sgn(nn') + \frac{1}{|n|} \right ) \dot{\zeta}_{n'}\zeta_{n - n'} + \frac{Q}{2 \pi R^{2}} \sum_{n' \neq 0} sgn(nn') \zeta_{n'}\zeta_{n - n'}. \nonumber \\ 
\end{eqnarray}
The overdot denotes total time derivative. The sign function $sgn(nn')=1$ if $(nn') > 0$ and $sgn(nn')=-1$ if $(nn') < 0$. We can use 
relations~(\ref{phi1t}) and~(\ref{phi2t}) to replace the 
velocity potentials $\phi_{i}$ in Darcy's law~(\ref{dimensionless2}) with 
the perturbation $\zeta$ and its time derivative $\dot{\zeta}$.
Keeping only quadratic terms in the perturbation 
amplitude, and equating Fourier modes $n$ on each side of Darcy's law, 
leads to the differential equation for perturbation 
amplitudes $\zeta_{n}$. For $n \neq 0$,
\begin{equation}
\label{result}
\dot{\zeta}_{n}=\lambda(n) \zeta_{n} + \sum_{n' \neq 0} \left [ F(n, n') \zeta_{n'} \zeta_{n - n'} + G(n, n') \dot{\zeta}_{n'} \zeta_{n - n'} \right ],
\end{equation}
where
\begin{equation}
\label{growth}
\lambda(n)= \left [ \frac{Q}{2 \pi R^{2}} \left (A |n| - 1 \right ) - \frac{\alpha}{R^{3}} |n| (n^{2} - 1) \right ],
\end{equation}
and
\begin{equation}
\label{F}
F(n, n')=\frac{|n|}{R} \left \{ \frac{QA}{2 \pi R^{2}} \left [ \frac{1}{2} - sgn(nn') \right ] - \frac{\alpha}{R^{3}} \left [ 1 - \frac{n'}{2} ( 3 n' + n )  \right ] \right \},
\end{equation}
\begin{equation}
\label{G}
G(n, n')=\frac{1}{R} \left \{ A|n| [ 1 - sgn(nn') ] - 1 \right \}.
\end{equation}

Equation~(\ref{result}) is the mode coupling 
equation of the Saffman-Taylor problem in a radial 
flow geometry Hele-Shaw cell. 
It gives us the time 
evolution of the perturbation amplitudes $\zeta_{n}$ accurate to 
second order. The first term on the 
right-hand side of equation~(\ref{result}) reproduces 
the linear stability analysis~\cite{{Pat},{Bat},{Wil}}, where 
$\lambda(n)$ denotes the linear growth rate.
The remaining terms represent second-order mode coupling 
in a radial flow geometry. 
They arise as a direct consequence of our weakly nonlinear analysis, which 
considers the presence of a full spectrum of modes.

It is interesting to compare the radial geometry mode coupling equation 
with the rectangular geometry result~\cite{Mir}. Several terms simplify. 
In particular,
\begin{equation}
\label{simplelambda}
\lambda(n) \rightarrow |k| \left [ A v_{\infty} - \alpha k^{2} \right ]
\end{equation}
\begin{equation}
\label{simpleF}
F(n, n') \rightarrow 0
\end{equation}
\begin{equation}
\label{simpleG}
G(n, n') \rightarrow A |k| \left [ 1 - sgn(kk') \right ].
\end{equation}
When $k$ and $k'$ are parallel there is no second order mode coupling 
in the rectangular geometry limit. Note that, 
in contrast to the radial geometry 
expression~(\ref{growth}), the rectangular geometry linear growth 
rate~(\ref{simplelambda}) is not time dependent. 
The approach of each term to its 
limit is ${\cal O}(1/R)$. In the following section we investigate the 
radial geometry equation~(\ref{result}) in more detail.

\section{Discussion}
\subsection{Linear evolution}

Considering the complexity of the flow, we begin by analyzing certain 
aspects of the linear theory. This analysis is needed for our 
subsequent weakly nonlinear investigation (section 3.2). 
First consider the purely linear 
contribution, which appears as 
the first term on the right hand side of equation~(\ref{result}). 
Since $R$ varies with time, the linear growth rate $\lambda(n)$ is time 
dependent as well. This implies that the actual relaxation or growth of 
mode $n$ is not proportional to the factor $\exp[\lambda(n)t]$, 
but rather 
\begin{equation}
\label{relax}
\zeta_{n}(t)=\zeta_{n}(0) \exp \left [ \int_{0}^{t} \lambda(n) dt' \right].
\end{equation}
If $\int_{0}^{t} \lambda(n) dt' > 0$ the disturbance 
grows, indicating instability. Inspecting equation~(\ref{growth}) for 
$\lambda(n)$ we notice opposing 
effects of the viscosity difference between the fluids 
(destabilizing) and of the surface tension at the interface 
(stabilizing). Two other relevant facts can be extracted from the 
linear growth rate: (i) the existence of a series of critical radii at 
which the interface becomes unstable for a given mode $n$ 
(defined by setting $\lambda(n)=0$), 
\begin{equation}
\label{critical}
R_{c}(n)=\frac{2 \pi \alpha}{Q} \frac{|n|(n^{2} - 1)}{(A|n| - 1)};
\end{equation}
(ii) the presence of a fastest growing mode $n^{\ast}$, given by the closest 
integer to the maximum of equation~(\ref{growth}) with respect to 
$n$ (defined by setting $d\lambda(n)/dn=0$), 
\begin{equation}
\label{fastest}
n_{max}(R)=\pm \sqrt{\frac{1}{3} \left (\frac{Q R A}{2 \pi \alpha} + 1 \right ) }.
\end{equation}
Note that $n^{\ast}$ varies with time. In view of equation~(\ref{relax}) 
$n^{\ast}$ is not simply related to the number of 
fingers present, even in the early stages of pattern formation. 
Furthermore, in the nonlinear regime the subsequent tip-splitting 
process and mode competition result in a final number of fingers which 
can differ from the number present in the linear regime.

Inspired by a model developed by Cardoso and Woods~\cite{Car} (their 
``model B''), we describe 
the behavior of the system assuming the presence of a constant low level 
of noise during the whole evolution of the interface. The sources of noise may 
come, for instance, from inhomogeneities on the surface of the Hele-Shaw 
cell, irregularities in the gap thickness $b$, or even from thermal or 
pressure fluctuations~\cite{Gin}. The predictions of this 
model are in qualitative agreement with experimental observations 
within the linear regime~\cite{Car}. We perform 
a detailed investigation of this model in the linear 
regime and extend its range of applicability to the weakly nonlinear 
stage of evolution.

Suppose that we 
begin with an initially circular interface that is steadily expanding. 
During the interface expansion each mode $n$ is perturbed with 
a constant (in time) random complex amplitude $\zeta_{n}(0)$. 
This noise amplitude contains an 
$n$ dependent random phase but its magnitude $|\zeta_{n}(0)|$ 
is independent of $n$ by assumption. As the interface continues 
to expand, it progressively reaches critical radii $R_{c}(n)$ for 
$n=$ $2$ , $3$ , $...$ . 
Once a particular $R_{c}(n)$ is reached, the 
perturbation amplitude $\zeta_{n}$ starts to vary with time. 
Within this model, the first 
order (linear) solution of equation~(\ref{result}) can be written as
\begin{equation}
\label{linear}
\zeta_{n}^{lin}(t)=\left\{ \begin{array}{cl}
\zeta_{n}(0) &\mbox{if $R < R_{c}(n)$} \\
\zeta_{n}(0) \left \{ \left ( \frac{R}{R_{c}(n)} \right )^{A|n| - 1} \exp \left [ (A|n| - 1) \left ( \frac{R_{c}(n)}{R} - 1 \right ) \right ] \right \} &\mbox{if $R \geq R_{c}(n)$}.
\end{array}\right.
\end{equation}

To further analyze the behavior of the linear solution 
equation~(\ref{linear}) and see the overall effect of the above mentioned 
properties, we plot the time evolution of the interface. 
It is convenient 
to rewrite the net perturbation~(\ref{z}) in terms of sine and cosine modes
\begin{equation}
\label{sincos}
\zeta(\theta,t)= \zeta_{0} + \sum_{n = 1}^{\infty} \left[ a_{n}(t)\cos(n\theta) + b_{n}(t)\sin(n\theta) \right ],
\end{equation}
where $a_{n}=\zeta_{n} + \zeta_{-n}$ and $b_{n}=i \left ( \zeta_{n} - \zeta_{-n} \right )$ are real-valued.
Throughout this work we use the experimental 
parameters given in Paterson's classical experiment~\cite{Pat}. 
Paterson observed the rapid growth of fingers, 
as air ($\eta_{1} \approx 0$) 
was blown at a relatively high injection rate, $Q=9.3$ $\rm{cm^{2}/s}$, 
into glycerine ($\eta_{2} \approx 5.21$  $\rm{g}$/($\rm{cm}$ $\rm{s}$)) 
in a radial source flow Hele-Shaw cell. The thickness of the cell 
$b=0.15$ $\rm{cm}$ and the surface tension $\sigma=63$ $\rm{dyne/cm}$. 
We take into account modes $n$ ranging from $n=2$ up to $20$. 
We evolve from initial radius $R_{0}=0.05$ $\rm{cm}$. The noise 
amplitude $|\zeta_{n}(0)|=R_{0}/500$. 
Figure 2, depicts the evolution of the interface, 
for a random choice of phases, up to time $t=30$ $\rm{s}$.

Linear evolution proceeds through a cascade of 
modes, with increasing participation 
of higher modes $n$ as time progresses. 
The number of fingers is typically given by the mode $n$ which has grown to 
largest amplitude. Variability of finger lengths depends 
primarily on modes of smaller $n$. Modes of larger $n$ 
can cause the tips of the fingers to split. 
For the early evolution of the interface, patterns like those shown 
in figure 2 resemble well the experimental patterns found 
in the literature~\cite{{Pat},{Che},{Tho},{Bat}}.

In contrast to the experiments, some important features are 
not present in figure 2, revealing failure of the 
purely linear approach. Spreading of the fingers, which is observed in 
the experiments, is not clearly shown in figure 2. 
Experimentally, the tips of fingers tend to split. To observe 
finger-splitting using the linear 
theory, we would have to go to unrealisticly long times. 
Finger competition, also, is not as pronounced in the linear theory 
as in experiment.
These last few remarks indicate that, despite being an adequate 
approximation for early times, 
a purely linear theory does not describe all the basic 
physical mechanisms (spreading, splitting and competition) 
involved in the pattern formation. The linear regime 
is valid only for a limited initial period beyond which nonlinear 
effects take over.

\subsection{Weakly nonlinear evolution}
The nonlinear effects considered here come from the contributions of the 
second term on the right hand side of equation~(\ref{result}). Let us 
begin by briefly describing the various pieces appearing in this 
second term. The function $F(n, n')$, which multiplies the term 
$ \zeta_{n'} \zeta_{n - n'} $, 
presents two parts (see equation~(\ref{F})) having distinct origins: the first 
one comes from the unperturbed radial 
flow (equation~(\ref{steady})) evaluated at the perturbed 
interface ${\cal R}$, while the second one arises 
from the second order terms in $\zeta$ present in the 
curvature (see equation~(\ref{kappa})). Neither term survives in the 
rectangular geometry limit. On the other hand, the function 
$G(n, n')$ (see equation~(\ref{G})) multiplies the term 
$ \dot{\zeta}_{n'} \zeta_{n - n'} $, which couples the perturbed flow 
$\dot{\zeta}$ with the interface shape perturbation $\zeta$. 
Only the ${\cal O}(1/R)$ part of $G(n, n')$ depends on the curvature of 
the unperturbed interface.

To visualize the implications of mode coupling, 
we solve equation~(\ref{result}) to second order accuracy. 
If we substitute the linear solution given in 
equation~(\ref{linear}) into the second-order terms on the right 
hand side of equation~(\ref{result}), we obtain the differential 
equation
\begin{equation}
\label{try2}
\dot{\zeta}_{n}=\lambda(n)\zeta_{n} + W(n,t),
\end{equation}
where 
\begin{equation}
\label{W}
W(n,t)=\sum_{n' \neq 0} \left [ F(n, n') \zeta_{n'}^{lin} \zeta_{n - n'}^{lin} + G(n, n') \dot{\zeta}_{n'}^{lin} \zeta_{n - n'}^{lin} \right ]
\end{equation}
acts as a driving force in the linearized equation of motion~(\ref{try2}).
Despite the complicated form of the functions $\lambda(n)$ and $W(n,t)$, 
equation~(\ref{try2}) is a standard first order linear differential 
equation~\cite{Gra} with the solution
\begin{equation}
\label{sol}
\zeta_{n}(t)=\left\{ \begin{array}{cl}
\zeta_{n}(0) &\mbox{if $R < R_{c}(n)$}\\
\zeta_{n}^{lin}(t) \left \{ 1 + \int_{t_{c(n)}}^{t} \left [ \frac{W(n,t')}{\zeta_{n}^{lin}(t')} \right ] dt' \right\} &\mbox{if $R \geq R_{c}(n)$}.
\end{array}\right.
\end{equation}
Here $t_{c}(n)$ is the time required for the unperturbed growth 
to reach radius $R_{c}(n)$ 
and can be easily calculated from equation~(\ref{R}).
This solution describes the weakly nonlinear evolution, where the dominant 
modes just become coupled by nonlinear effects.

We use the second order solution~(\ref{sol}) to investigate 
the nonlinear coupling among various modes $n$. 
Once again, we use the sine and 
cosine mode representation for the net perturbation $\zeta(\theta,t)$. 
In figure 3, we plot the 
interface for a certain time ($t=30$ ${\rm s}$), 
considering the same random choice of initial phases as was employed 
in figure 2, and coupling all modes with $2 \le n \le 20$. 
The solid curve represents the 
interface shape obtained from the mode coupling solution~(\ref{sol}), and 
the dashed line represents the interface in the linear 
approximation (taken from figure 2). 
The two interfaces are plotted together 
to facilitate comparison between the linear and nonlinear approaches.
As we can see, the nonlinear evolution leads to 
wider fingers and their tips become more blunt. These fingers spread 
and some of them start to bifurcate, by splitting at the tip.
The onset of finger competition is also observed, 
with some fingers slightly enhanced and others diminished. 

Even though the differences between the two evolutions are not
dramatic, we see that the nonlinear approach developed here leads to a
more realistic description of the radial fingering patterns. The
effects of finger spreading, finger tip-splitting and finger
competition are all present.  The fact that the second order pattern
is quantitatively close to the linear pattern suggests that the
nonlinearities are fairly accurately described up to $t=30$ ${\rm s}$
by second order mode coupling theory, because the contributions of
third and higher order terms should be smaller yet.

One interesting point to be discussed is the evolution of 
a particular mode and 
its interaction with its sub-harmonics and harmonics (figure 4). 
Here we concentrate our attention on a 
small number of modes. This situation is obviously idealized compared 
with interaction of all modes illustrated in figure 3, 
but it gives us good indications about the role of mode coupling 
during interface evolution.

For convenience we write the explicit equations of motion in the sine
and cosine representation of the interface.  The even and odd
functions $a_{n} \cos(n \theta)$ and $b_{n} \sin(n \theta)$ form
representations (respectively, the identity $A_{1}$, and the
reflection antisymmetric representation $A_{2}$) of the rotation and
reflection symmetry group $C_{nv}$~\cite{Lan}. The signs, and relative
magnitudes, of $a_{n}$ and $b_{n}$ determine the phase of mode $n$. In
real space, they determine the location of the $n$ fingers on the
growing perimeter.  In addition to these one-dimensional
representations, $C_{nv}$ has a set of two dimensional irreducible
representations $\{E_m\}$ formed by the cosine and sine functions of
mode $m=1,2,\cdots$, up to the largest integer less than $n/2$. Modes
$m^{\prime}>n/2$ transform identically to modes $m=n-m^{\prime}<n/2$.
When $n$ is even, $m=m^{\prime}=n/2$ is an integer, and the
representation $E_{n/2}$ decomposes into a pair of reflection
symmetric and antisymmetric representations $B_1$ and $B_2$.

The transformation properties of modes under rotations and reflections
guarantee invariance of their equations of motion, but constrain the
forms of those equations. Symmetry dictates which products of modes
appear in each equation of motion. The numerical values of the
coefficients multiplying these products, however, can only be obtained
from a full derivation of the equation of motion such as we presented
in section~\ref{derivation}. Symmetry issues have been addressed 
in studies of shape selection for Saffman-Taylor fingers 
in rectangular channels~\cite{{Tan2},{Com}} 
and crystal growth processes~\cite{{Bre},{Kup},{Ama}}.

First, we consider the influence of a fundamental mode on the growth 
of its harmonic. We take $n$ as the fundamental and $2n$ as the 
harmonic. Without loss of generality we may take the phase of 
the fundamental so that $a_{n} > 0$ and $b_{n}=0$. We also replace 
$\dot{a}_{n}$ with $\lambda(n)a_{n}$ in terms already of second order. 
The equations of motion become
\begin{equation}
\label{new3}
\dot{a}_{2n}=\lambda(2n)a_{2n} + \frac{1}{2} \left [ F(2n, n) + \lambda(n) G(2n, n) \right ]a_{n}^2
\end{equation}
\begin{equation}
\label{new4}
\dot{b}_{2n}=\lambda(2n)b_{2n}.
\end{equation}
To see how symmetries constrain the form of these equations, we employ
the group $C_{2nv}$. The fundamental $a_n$ transforms as
representation $B_1$, while the harmonic $a_{2n}$ transforms as the
identity representation $A_1$. The product of $B_1$ with itself
equals the identity representation, so a term of the form $a_n^2$
appears in equation~(\ref{new3}). In contrast, $b_{2n}$ transforms as
the reflection antisymmetric representation $A_2$, which cannot be
formed from a product of the representation $B_1$ with any other
representation present. Consequently $a_n$ cannot influence the growth
of $b_{2n}$ at second order.

Note that $(1/2) \left [ F(2n, n) + \lambda(n) G(2n, n) \right ] < 0$.
At second order, the result is a driving term of order $a_{n}^{2}$
forcing growth of $a_{2n} < 0$. With this particular phase of the
harmonic forced by the dynamics, the $n$ fingers of the fundamental
mode $n$ tend, first, to spread and blunt, and later, to split. See
figure 4 for an illustration. The solid line exhibits the broadened
and split fingers predicted by the nonlinear mode coupling, while the
dashed line shows the linear theory for comparison. Had growth of
$a_{2n} > 0$ been favored, fingers would have sharpened instead of
blunting. Mode $b_{2n}$, whose growth is uninfluenced by $a_{n}$,
skews the fingers of mode $n$. In the presence of $a_{2n} < 0$, the
role of $b_{2n}$ is to favor one of the two split fingers over the
other. The driving term in equation of motion~(\ref{new3})
spontaneously generates the harmonic mode, even in the absence of
random noise. This effect is clearly shown in the numerical solution
of a solidification model~\cite{Brush2}.

In the rectangular geometry limit the driving term vanishes. Finger
splitting thus depends on the curvature of the unperturbed interface.
Finger spreading and splitting is a unique prediction of the second
order mode coupling in the radial geometry.

Next, consider the influence of a fundamental mode $n$, assuming $n$
is even, on the growth of its sub-harmonic mode $n/2$. Again, without
loss of generality, we may choose the phase for the fundamental to
set $a_{n} > 0$ and $b_{n}=0$. The equations of motion become
\begin{equation}
\label{new5}
\dot{a}_{n/2}=\left \{ \lambda(n/2) + \frac{1}{2} \left [ F \left( -\frac{n}{2}, \frac{n}{2} \right ) + \lambda(n/2) G \left ( \frac{n}{2}, -\frac{n}{2} \right ) \right ]a_{n} \right \}a_{n/2}
\end{equation}
\begin{equation}
\label{new6}
\dot{b}_{n/2}=\left \{ \lambda(n/2) - \frac{1}{2} \left [ F \left ( -\frac{n}{2}, \frac{n}{2} \right ) + \lambda(n/2) G \left (\frac{n}{2}, -\frac{n}{2} \right ) \right ]a_{n} \right \}b_{n/2}.
\end{equation}
As in our analysis of the growth of the harmonic
(equations~(\ref{new3}) and~(\ref{new4})) the multiplication of
representations dictates the form of the equations of motion. We
employ the symmetry group $C_{nv}$. Mode $n$ forms the identity
representation $A_1$. Note that $a_{n/2}$ and $b_{n/2}$ transform as
representations $B_1$ and $B_2$. Had we taken $n$ odd, there would be no
sub-harmonic $n/2$. However, the cosine and sine modes $m$ transform as
two-dimensional representations $E_m$ for $m<n/2$ and $E_{n-m^{\prime}}$ for
$m^{\prime}>n/2$. Consider a pair of modes $m$ and $m^{\prime}=n-m$. The
product of the fundamental mode $n$ with mode $m^{\prime}$ transforms
as mode $m$. Consequently mode $n$ and $m^{\prime}$ together would
influence the growth of mode $m$. The special case under consideration
has $n$ even, and $m=m^{\prime}=n/2$.

Note that $(1/2) \left [ F(-n/2, n/2) + \lambda(n/2) G(n/2, -n/2)
\right ] > 0$.  The fundamental thus enhances the growth of its even
sub-harmonic while inhibiting growth of its odd sub-harmonic. The
result is to favor perturbations $a_{n/2}$ which break the $n$-fold
rotational symmetry of the fundamental by alternately increasing and
decreasing the length of each of the $n$ fingers.  This effect
describes finger competition.  In contrast, mode $b_{n/2}$ skews the
fingers while varying the depths of the valleys between fingers.
Whereas the equations of motion for the
harmonic~(equations~(\ref{new3}) and~(\ref{new4})) contain a driving
term, the equations of motion for the
sub-harmonic~(equations~(\ref{new5}) and~(\ref{new6})) show that 
rotational symmetry breaking sub-harmonics must be introduced through 
noise or initial conditions. They cannot grow spontaneously. The sign of 
$a_{n/2}$ is not determined by the equation of motion.  Positive and
negative values of $a_{n/2}$ determine which fingers grow at the
expense of the others.  Finger competition with $a_{n/2} > 0$ is
evident in figure 4. Reflection-symmetry breaking perturbations 
$b_{n/2}$ are suppressed consistent with known properties of selected 
steady state patterns~\cite{{Tan2},{Com}}.

Finally, consider the influence of a mode on itself. Consider $a_{n} >
0$ and $b_{n}= 0$. There is no ${\cal O}(a_{n}^{2})$ term in the
equation of motion, consistent with the demands of $C_{\infty v}$
invariance of the radial geometry equations of motion. However, our
discussion above shows that a fundamental mode $n$ spontaneously
generates its own harmonic $a_{2n} < 0$ of order $a_{n}^2$.  This
harmonic couples back into the equation of motion for $a_{n}$ (replace
$n/2 \rightarrow n$ in equation~(\ref{new5})), resulting in an ${\cal
O}(a_{n}^{3})$ diminution over the linear growth of $a_{n}$. In real
space, a growing finger broadens as it pushes into the more viscous
fluid. The broadened finger becomes increasingly resistant to further
growth. The third order coupling of a mode with itself was studied in
related solidification models~\cite{Brush,Debroy}.

In order to reinforce the conclusions of the previous paragraphs, 
compare the time evolution of the fundamental, harmonic and 
sub-harmonic perturbation amplitudes following 
the linear~(\ref{linear}) and nonlinear~(\ref{sol}) solutions. In figure 5 
we present the time evolution of the perturbation amplitudes $a_{n}$ for 
the modes $n$=4, 8 and 16 as given by the simple linear superposition of these 
modes (dashed curves) and by our weakly nonlinear mode coupling theory 
(solid curves). 
We observe that the weakly nonlinear coupling dictates 
the enhanced growth of both sub-harmonic ($a_{4}$) and 
first harmonic ($a_{16}$) modes while diminishing the growth 
of the fundamental ($a_{8}$). The sign of the harmonic ($a_{16}$) 
is dictated as well, going strongly negative although its initial value 
was positive. Therefore, the nonlinear effects 
naturally enhance finger competition and tendency to finger tip-splitting.

\section{Concluding remarks}

In this paper, we studied the onset of nonlinear effects in the radial
viscous fingering in a Hele-Shaw cell.  While reasonably
accurate for the initial stages of the flow, linear stability theories
do not explain well the origin of the essential mechanisms of finger
spreading, finger tip-splitting and finger competition.  We derived a
second order mode coupling differential equation for the perturbation
amplitudes.  By considering the interaction of a small number of
modes, we revealed the role of sub-harmonic and harmonic perturbations
at the onset of pattern formation. Symmetries of the radial flow
geometry constrain the form of the mode coupling equations. Mode 
coupling provides a mechanism for finger spreading, splitting and competition
creating patterns consistent with those observed 
in experiments. Reflection-symmetric fingers are favored. Our analytical mode
coupling approach links the simple initial
linear behavior with the strongly nonlinear advanced stages of
evolution.

We compared the radial flow geometry with the simpler case of 
flow in rectangular channels~\cite{Mir}. The rectangular geometry 
represents a special case of the radial flow, in which effects of curvature 
of the unperturbed interface vanish. Finger splitting, in the radial 
flow geometry, is driven by curvature of the unperturbed interface. 
Tendency towards finger splitting may be controlled by adjusting this 
interfacial curvature. Presumably similar control is possible by 
performing the flow in curved two-dimensional spaces. Such effectively 
two-dimensional curved spaces may be found between curved surfaces 
embedded in three dimensions, separated from each other by a small 
distance $b$. Positive spatial curvature 
(e.g., flow between concentric spheres) should inhibit finger splitting 
while negative curvature (e.g., flow between saddle shape surfaces) 
should enhance it.

Finally, the weakly nonlinear 
approach we developed in this paper and 
in reference~\cite{Mir} will be useful to investigate the onset of nonlinear 
effects in hydrodynamic stability problems involving 
ferrofluids~\cite{{Ros},{Jac}} and lipid domains~\cite{Sto}.

\vskip 0.5 in
\noindent
{\bf Acknowledgments}\\
\noindent
This work was supported in part by the National Science Foundation
grant No. DMR-9221596. J.A.M. (CNPq reference number 200204/93-9)
would like to thank CNPq (Brazilian Research Council) for financial
support.  We acknowledge useful discussions with Steve Garoff, David
Jasnow and Robert F. Sekerka.
\pagebreak

\pagebreak

\noindent
{\Large {\bf Figure Captions}}
\vskip 0.5 in
\noindent
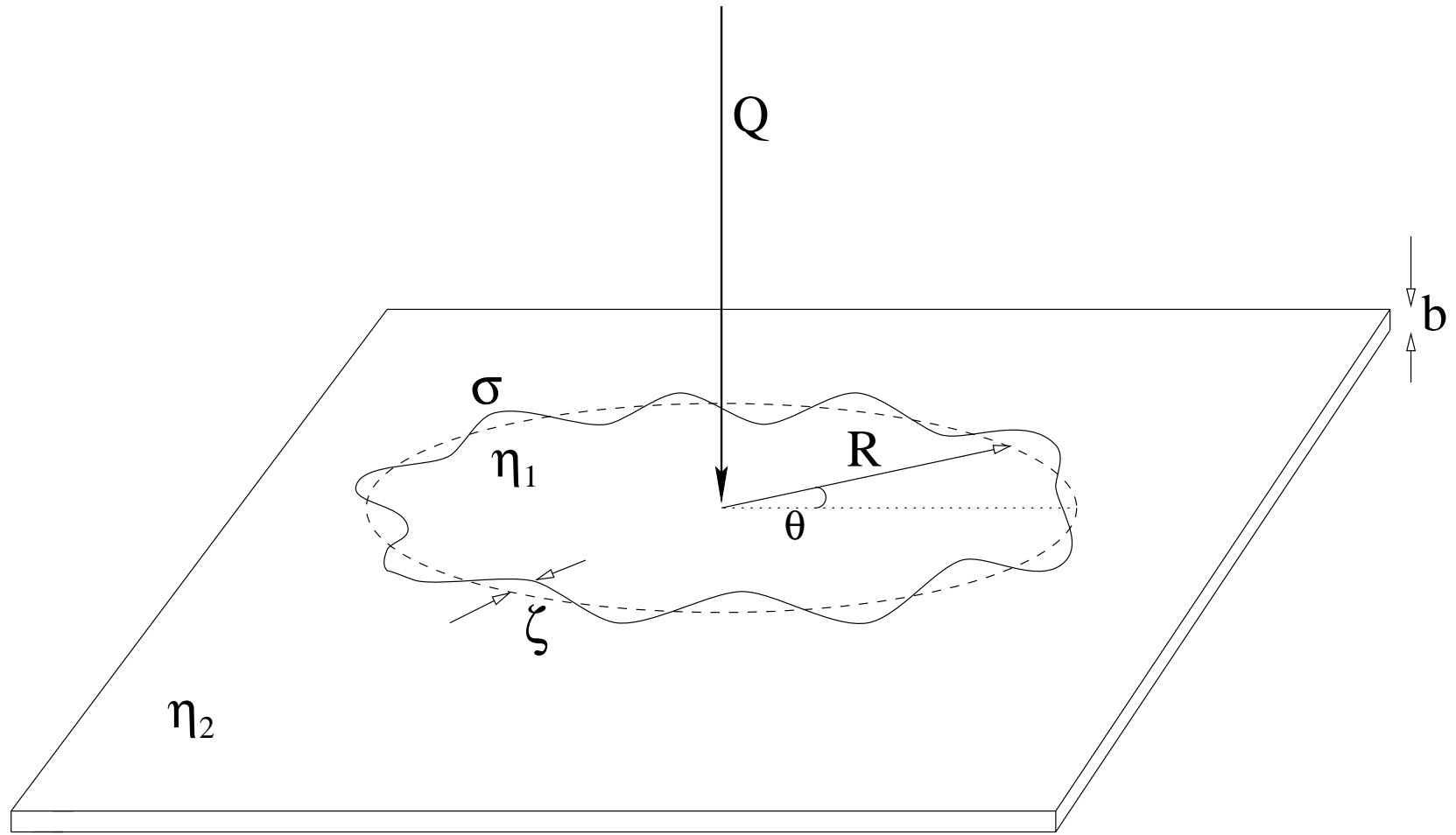
{\bf Figure 1:} Schematic configuration of the radial source flow. 
The viscosities of inner and outer fluids are 
$\eta_{1}$ and $\eta_{2}$, respectively. Fluid 1 is injected into the 
Hele-Shaw cell, previously filled with fluid 2, 
with constant injection rate $Q$. 
The dashed line represents the unperturbed circular interface 
$R$ and the solid 
undulated curve depicts the perturbed interface ${\cal R}= R + 
\zeta(\theta,t)$, where $\theta$ is the polar angle.
The surface tension between the two fluids is given by $\sigma$ 
and $b$ denotes the thickness of the Hele-Shaw cell.
\vskip 0.25 in
\noindent
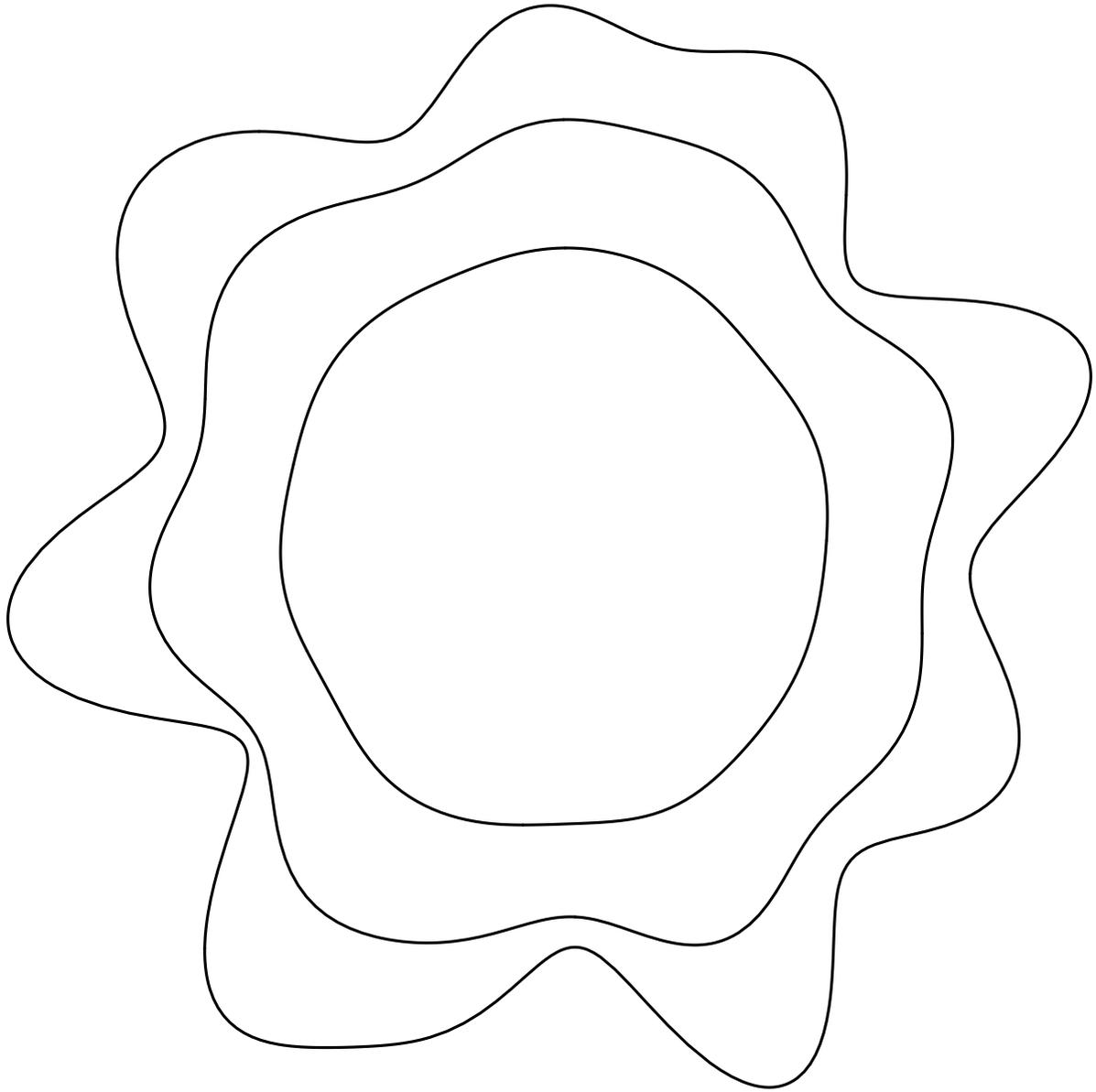
{\bf Figure 2:} Time evolution of the interface between the fluids, as 
given by the linear theory, 
including modes $2 \leq n \leq 20$. The initial perturbation 
amplitudes $|\zeta_{n}(0)|$=$R_{0}/500$ and $R_{0}=0.05$ ${\rm cm}$. 
The experimental parameters are given in the text and 
$t$=10, 20 and 30 ${\rm s}$.
\vskip 0.25 in
\noindent
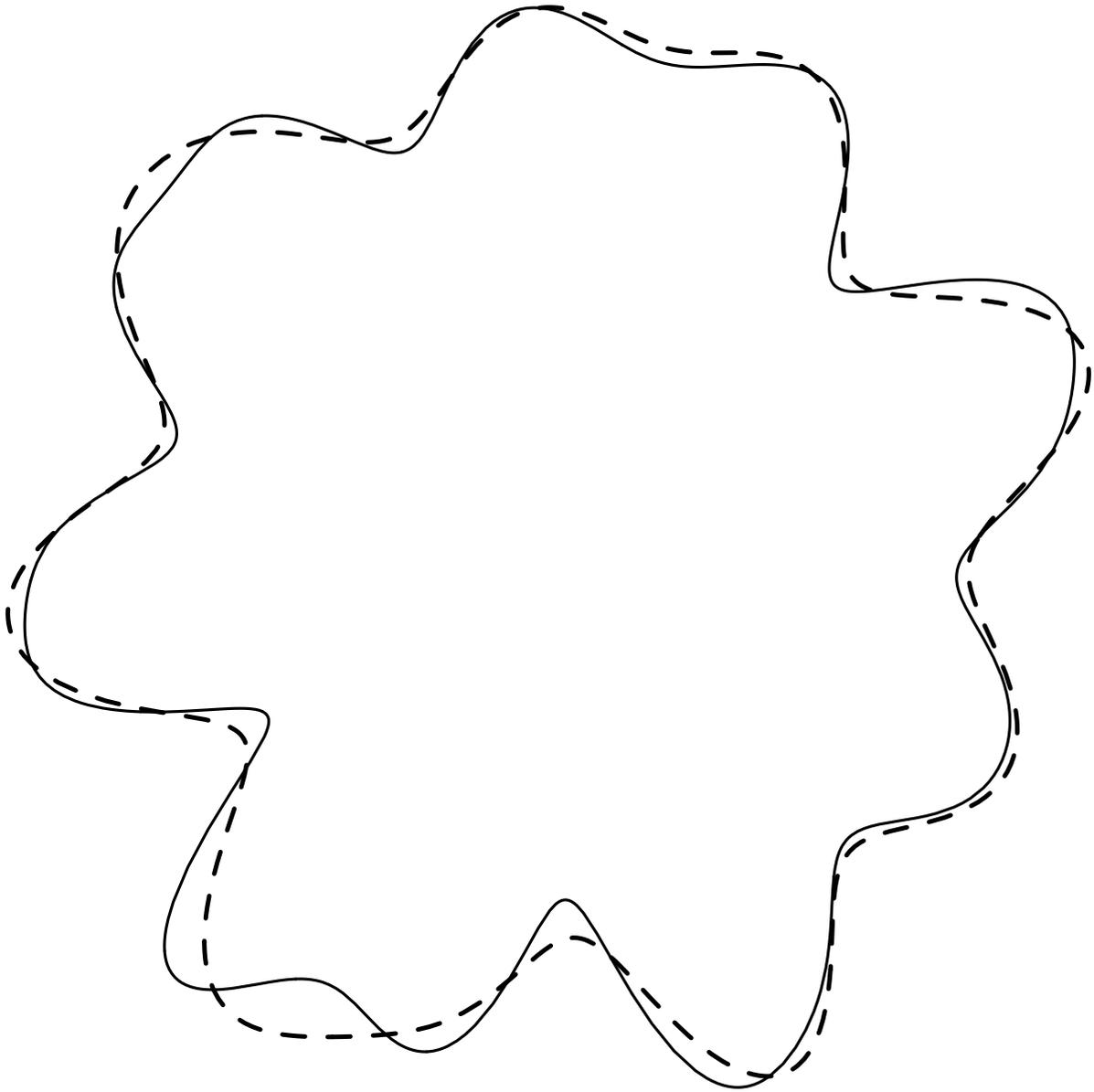
{\bf Figure 3:} Snapshot of the interfacial patterns for $t$=30 ${\rm s}$ 
in the cases of linear (dashed curve) and nonlinear (solid curve) evolutions.
All physical parameters and initial conditions are the same as those used in 
figure 2. 
\vskip 0.25 in
\noindent
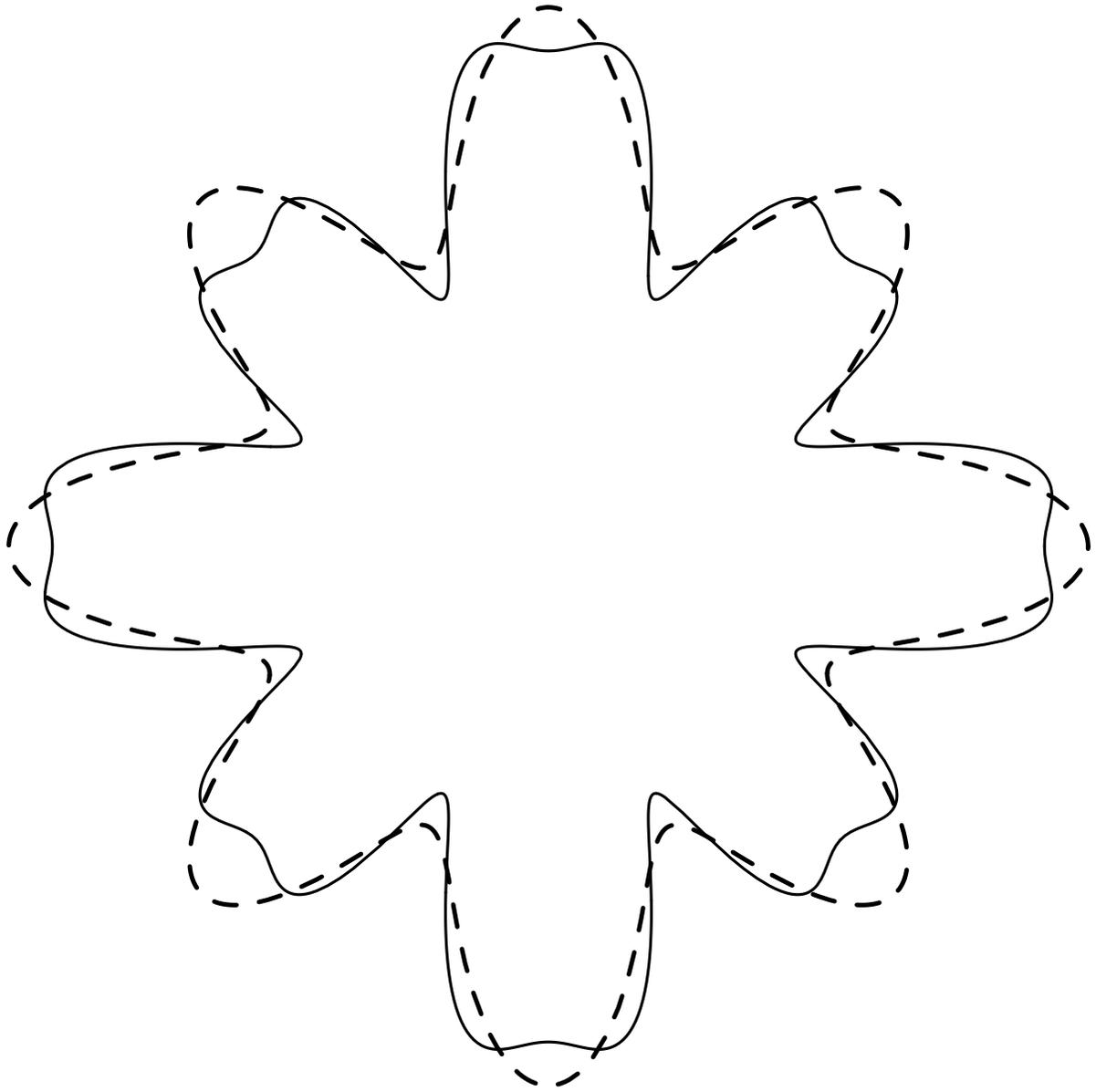
{\bf Figure 4:} Snapshot of the interface between the two fluids 
($t$=30 ${\rm s}$), for the case of three interacting cosine modes 
(a fundamental $n=8$, its sub-harmonic $n/2=4$ and its harmonic 
$2n=16$). Initial conditions are $a_{n}(0)$=$3.5 \times 10^{-4}$ ${\rm cm}$ 
and $R_{0}=0.3$ ${\rm cm}$. 
Other parameters are as in figures 2 and 3. 
The dashed (solid) curve obeys the linear (nonlinear) theory. 
The nonlinear contributions coming from the harmonic results 
in wider fingers, which tend to bifurcate. 
Finger competition is favored by the influence of the 
sub-harmonic mode.
\vskip 0.25 in
\noindent
\vskip 0.25 in
\noindent
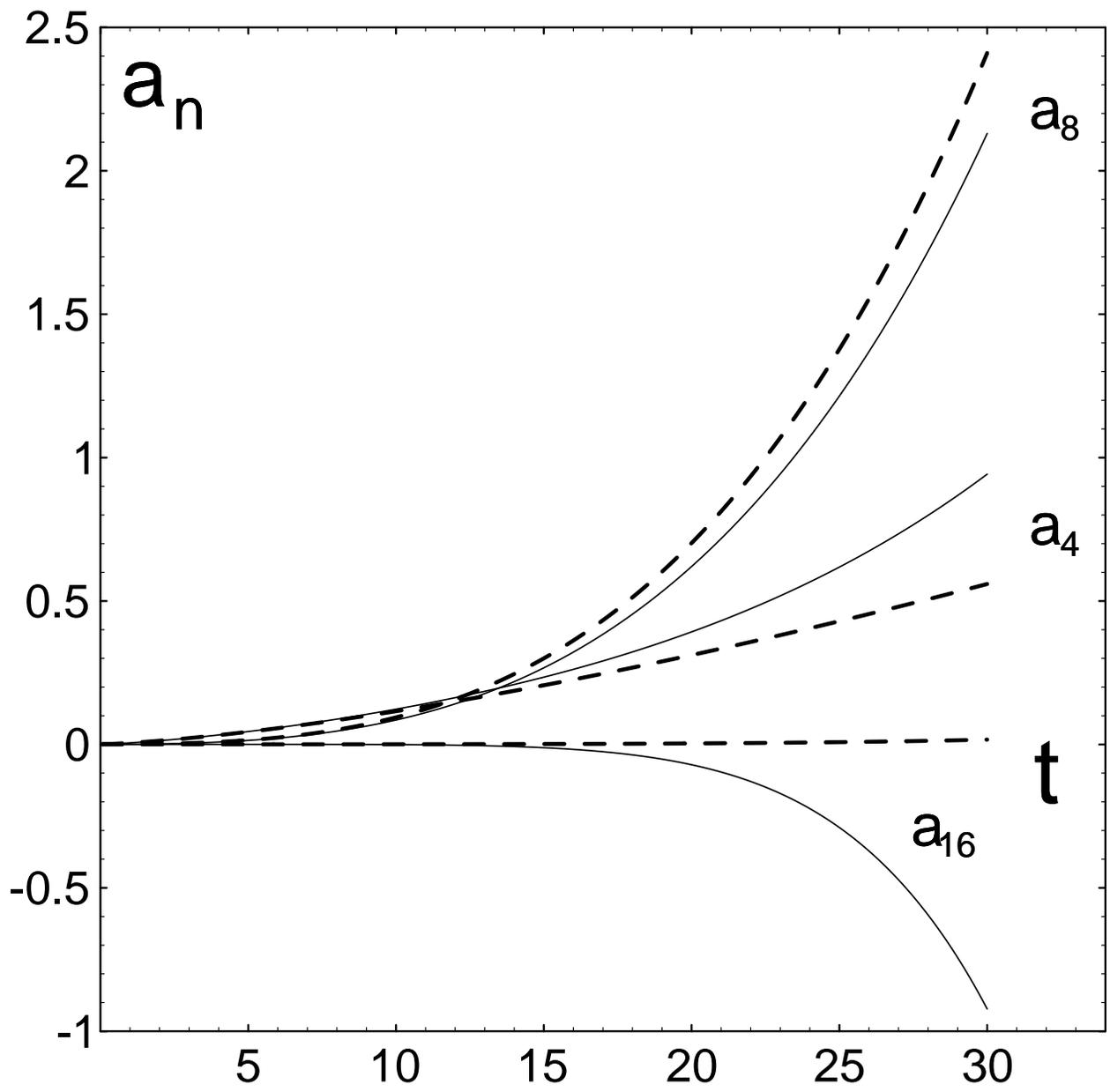
{\bf Figure 5:} Time evolution of the cosine perturbation 
amplitudes $a_{n}$ ($n$=4, 8 and 16) for linear (dashed curves) 
and nonlinear (solid curves) cases. 
The initial conditions and physical parameters are the same as 
those used in figure 4.
\vskip 0.25 in
\noindent
\vskip 0.25 in
\noindent

\begin{thebibliography}{99}


\bibitem{Rev}For review articles on this subject, see for instance, D. Bensimon, L. P. Kadanoff, S. Liang, B. I. Shraiman and C. Tang, Rev. Mod. Phys. {\bf 58}, 977 (1986); S. Tanveer, in {\it Asymptotics beyond all orders}, 1991, NATO ASI series B, vol. {\bf 284}, p. 131, edited by H. Segur, S. Tanveer and H. Levine (Plenum, New York, 1991); K. V. McCloud and J. V. Maher, Phys. Rep. {\bf 260}, 139 (1995). See also reference~\cite{Hom}. 

\bibitem{BenJac-2}E. Ben-Jacob, Nature {\bf 343}, 523 (1990).

\bibitem{Saf}P. G. Saffman and G. I. Taylor, Proc. R. Soc. London Ser. A {\bf 245}, 312 (1958).

\bibitem{Pat}L. Paterson, J. Fluid Mech. {\bf 113}, 513 (1981).

\bibitem{Tan}M. Siegel, S. Tanveer and W.-S. Dai, J. Fluid Mech. {\bf 323}, 201 (1996) and references therein; S. Tanveer, Phil. Trans. R. Soc. Lond. A {\bf 343}, 155 (1993).  

\bibitem{Hou}T. Y. Hou, J. S. Lowengrub and M. J. Shelley, J. Comp. Phys. {\bf 114}, 312 (1994); W.-S. Dai and M. J. Shelley, Phys. Fluids A {\bf 5}, 2131 (1993). 

\bibitem{Brush2}L. N. Brush and R. F. Sekerka, J. Crystal Growth {\bf 96}, 419 (1989).

\bibitem{Brush}L. N. Brush, R. F. Sekerka and G. B. McFadden, J. Crystal Growth {\bf 100}, 89 (1990).

\bibitem{Debroy}P. P. Debroy and R. F. Sekerka, Phys. Rev. E {\bf 53}, 6244 (1996).

\bibitem{Mir}J. A. Miranda and M. Widom (preprint, 1997).

\bibitem{Che}J.-D. Chen, J. Fluid Mech. {\bf 201}, 223 (1989); J. -D. Chen, Exp. Fluids {\bf 5}, 363 (1987).

\bibitem{Tho}H. Thom\'e, M. Rabaud, V. Hakim and Y. Couder, Phys. Fluids {\bf A1}, 224 (1989).

\bibitem{Bat}J. Bataille, Revue Inst. P\'etrole {\bf 23}, 1349 (1968).

\bibitem{Hom}G. Homsy, Ann. Rev. Fluid Mech. {\bf 19}, 271 (1987).

\bibitem{Wil}S. D. R. Wilson, J. Colloid Interface Sci. {\bf 51}, 532 (1975).

\bibitem{Sad}J. E. Sader, D. Y. C. Chan, and B. D. Hughes, Phys. Rev. E 
{\bf 49}, 420 (1994).

\bibitem{Car2}Ll. Carrillo, F. X. Magdaleno, J. Casademunt and J. Ort\'{\i}n, 
Phys. Rev. E {\bf 54}, 6260 (1996).

\bibitem{Car}S. S. S. Cardoso and A. W. Woods, J. Fluid Mech. {\bf 289}, 351 (1995).

\bibitem{Dub}B. A. Dubrovin, A. T. Fomenko, and S. P. Novikov, {\it Modern Geometry-Methods and Applications, Part 1} (Springer-Verlag, New York, 1984).

\bibitem{McL}J. W. McLean and P. G. Saffman, J. Fluid Mech. {\bf 102}, 455 (1981); P. G. Saffman, in {\it Macroscopic Properties of Disordered Media}, 1982, Lecture Notes in Physics, vol. {\bf 154}, p. 208, edited by R. Burridge, S. Childress and G. Papanicolaou (Springer-Verlag, New York, 1982).

\bibitem{Par}For a discussion on the influence of wall wetting effects 
on the curvature in the direction perpendicular to the plates, see C. -W. Park and G. M. Homsy, J. Fluid Mech. {\bf 139}, 291 (1984); D. A. Reinelt, J. Fluid Mech. {\bf 183}, 219 (1987). 

\bibitem{Ros}R. E. Rosensweig, {\it Ferrohydrodynamics} (Cambridge University Press, Cambridge, 1985).

\bibitem{Gin}M. J. P. Gingras and Z. R\'acz, Phys. Rev. A {\bf 40}, 5960 (1989).

\bibitem{Gra}I. S. Gradshteyn and I. M. Ryzhik, {\it Table of Integrals, Series, and Products} (Academic Press, New York, 1994).

\bibitem{Lan} L. D. Landau and E. M. Lifshitz, {\it Quantum Mechanics: Non-relativistic Theory} (Pergamon Press, New York, 1977).

\bibitem{Tan2}S. Tanveer, Phys. Fluids {\bf 30}, 1589 (1987).

\bibitem{Com}R. Combescot and T. Dombre, Phys. Rev. A {\bf 38}, 2573 (1988).

\bibitem{Bre}E. Brener, H. M\"{u}ller-Krumbhaar, Y. Saito and D. Temkin, Phys. Rev. E {\bf 47}, 1151 (1993); T. Ihle and H. M\"{u}ller-Krumbhaa, Phys. Rev. E {\bf 49}, 2972 (1994). 

\bibitem{Kup}R. Kupferman, D. A. Kessler and E. Ben-Jacob, Physica A {\bf 213}, 451 (1995).

\bibitem{Ama}M. B. Amar and E. Brener, Physica D {\bf 98}, 128 (1996).

\bibitem{Jac}D. P. Jackson, R. E. Goldstein and A. O. Cebers, Phys. Rev. E {\bf 50}, 298 (1994) and references therein.

\bibitem{Sto}H. A. Stone and H. M. McConnell, Proc. R. Soc. London Ser. A {\bf 448}, 97 (1995) and references therein.

\end{thebibliography}
\end{document}